\documentclass[3p,times,twocolumn]{elsarticle}
\usepackage{ifthen}
\usepackage{xspace}
\usepackage{booktabs}
\usepackage{multirow}
\usepackage{ctable}
\usepackage{calc}
\usepackage{relsize}
\usepackage[colorlinks=false,pdfborder={0 0 0}]{hyperref}

\newcommand{\format}{1}
\newcommand{\ifdraft}[2]{\ifthenelse{\equal{\format}{0}}{#1}{#2}}
\newcommand{\ifarxiv}[2]{\ifthenelse{\equal{\format}{1}}{#1}{#2}}
\newcommand{\ifnproc}[2]{\ifthenelse{\equal{\format}{2}}{#1}{#2}}

\ifarxiv{
  \usepackage{amsmath}
  \usepackage{amssymb}
  \journal{Nuclear Physics B Proceedings}
}{
  \usepackage{ecrc}
  \usepackage{amsmath}
  \usepackage{amssymb}
  \journalname{Nuclear Physics B Proceedings Supplement}
  \jnltitlelogo{Nuclear Physics B Proceedings Supplement}
  \jid{nuphbp}
  \volume{00}
  \firstpage{1}
  \runauth{P. Ilten}
}

\ifdraft{
  \usepackage[pdf,clean]{svg}
  \usepackage{lineno}
  \linenumbers
}{}

\newcommand{\mn}{\ensuremath{\nu_\mu}\xspace}
\newcommand{\tn}{\ensuremath{\nu_\tau}\xspace}

\newcommand{\lep}{\ensuremath{\ell}\xspace}
\newcommand{\lepn}{\ensuremath{\nu_\ell}\xspace}
\newcommand{\had}{\ensuremath{h}\xspace}
\newcommand{\z}{\ensuremath{Z}\xspace}
\newcommand{\w}{\ensuremath{W}\xspace}
\newcommand{\ditau}{\ensuremath{{\tau\tau}}\xspace}
\newcommand{\mumu}{\ensuremath{\tau_\mu\tau_\mu}\xspace}
\newcommand{\mue}{\ensuremath{\tau_\mu\tau_e}\xspace}
\newcommand{\emu}{\ensuremath{\tau_e\tau_\mu}\xspace}
\newcommand{\muh}{\ensuremath{\tau_\mu\tau_\had}\xspace}
\newcommand{\eh}{\ensuremath{\tau_e\tau_\had}\xspace}

\newcommand{\pt}{\ensuremath{p_\mathrm{T}}\xspace}
\newcommand{\kt}{\ensuremath{k_\mathrm{T}}\xspace}
\newcommand{\et}{\ensuremath{E_\mathrm{T}}\xspace}

\newcommand{\met}{\ensuremath{{{\not\mathrel{E}}_T}}\xspace}

\newcommand{\nb}{\ensuremath{\mathrm{nb}}\xspace}

\newcommand{\pb}{\ensuremath{\mathrm{pb}}\xspace}
\newcommand{\ipb}{\ensuremath{\mathrm{pb}^{-1}}\xspace}

\newcommand{\ifb}{\ensuremath{\mathrm{fb}^{-1}}\xspace}
\newcommand{\tev}{\ensuremath{\mathrm{TeV}}\xspace}
\newcommand{\gev}{\ensuremath{\mathrm{GeV}}\xspace}

\newcommand{\lhc}{LHC\xspace}
\newcommand{\lhcb}{LHCb\xspace}
\newcommand{\atlas}{ATLAS\xspace}
\newcommand{\cms}{CMS\xspace}
\newcommand{\fewz}{{\sc Fewz}\xspace}
\newcommand{\dynnlo}{{\sc Dynnlo}\xspace}
\newcommand{\lum}{\ensuremath{\mathcal{L}}\xspace}
\newcommand{\tls}{\ensuremath{\tau}-leptons\xspace}
\newcommand{\tl}{\ensuremath{\tau}-lepton\xspace}

\newcolumntype{R}{>{$}r<{$}}
\newcolumntype{L}{>{$}l<{$}}
\newcolumntype{C}{>{$}c<{$}}
\newcolumntype{E}{@{\extracolsep{\length}}>{$\pm}c<{$}@{\extracolsep{\length}}}

\begin{document}

\begin{frontmatter}
  
  % Title and author.
  \title{$Z \to \ditau$ and $W \to \tau\tn$ Cross-Sections at the LHC}
  \author{Philip Ilten \\ on behalf of the \lhcb Collaboration \\
    including results from \atlas and \cms}
  \address{School of Physics, University College Dublin}
  
  % Abstract.
  \begin{abstract}
    Measurements of the $\z \to \ditau$ and $\w \to \tau\tn$
    cross-sections at the LHC with data taken at $\sqrt{s} = 7$ TeV
    are reported for the \atlas, \cms, and \lhcb experiments. All
    results are found to agree with the Standard Model.
  \end{abstract}
  
  % Keywords.

\end{frontmatter}

\section{Introduction}

The production of $Z$ and $W$ bosons from $pp$-collisions, and their
subsequent decays to \tls at the Large Hadron Collider (\lhc) not only
provides important tests of the Standard Model (SM), but also lays the
groundwork for the study of beyond the SM physics using \tl
signatures.

Decays of \z bosons to \tl pairs are both a mechanism for
experimentally measuring hadronic \tl identification
efficiencies~\cite{atlasNoteTau, cmsPaperTau} and a calibration
channel for neutral Higgs searches~\cite{atlasNoteHn, cmsPaperHn}. The
final states of \tls produced from \w bosons can be used to measure
the polarization of the \w~\cite{atlasPaperP} boson or to search for
charged Higgses~\cite{atlasPaperHc, cmsPaperHc}.

In this review, the complete set of $\z \to \ditau$ and $\w \to
\tau\tn$ cross-sections, as measured by the \atlas, \cms, and \lhcb
experiments on the LHC, are reported using $2010$ and $2011$ datasets
taken at $\sqrt{s} = 7 ~\tev$. A full summary of the results,
including corresponding references, is provided in
Table~\ref{tab:results}, and a comparison of the theoretical agreement
of the results is given in Figure~\ref{fig:comparison}.

All three detectors are fully instrumented with charged particle
trackers, electromagnetic and hadronic calorimeters, and muon
systems. Both \atlas~\cite{atlasDetector} and \cms~\cite{cmsDetector}
are general purpose detectors designed to cover a central
pseudorapidity range of $|\eta| < 2.4$, while the
\lhcb~\cite{lhcbDetector} detector is a forward arm spectrometer,
purpose built for $B$-hadron physics, covering the forward
pseudorapidity range $2.0 < \eta < 5.0$.

\begin{table*}[!htb]
  \small
  \begin{center}
    \newlength{\length}
    \settowidth{\length}{\space}
    \begin{tabular}{c|LL|RR|RERERER@{\extracolsep{\length}}l|L@{\extracolsep{\length}}lCc}
      
      % Header.
      \toprule
      \multicolumn{1}{c}{Exp.}
      & \multicolumn{2}{c}{Analysis}
      & \multicolumn{1}{c}{$N_\mathrm{obs}$}
      & \multicolumn{1}{c}{$N_\mathrm{bkg}$}
      & \multicolumn{8}{c}{$\sigma\pm\mathrm{stat.}\pm\mathrm{syst.}\pm\mathrm{lumi.}$}
      & \multicolumn{2}{c}{\lum} 
      & \multicolumn{1}{c}{Data} 
      & \multicolumn{1}{c}{Ref.} \\
      \midrule[\heavyrulewidth] 
      % ATLAS results.
      \multirow{5}{*}{\atlas} & \multirow{4}{*}{$\z \to$} & \mumu        & 90   & 47  & 0.96 && 0.22 && 0.12 && 0.03 & nb     & 36                    & \ipb                  & 2010                    & \cite{atlasPaperZ}                 \\
                              &                           & \mue         & 1035 & 56  & 0.96 && 0.03 && 0.09 && 0.04 & nb     & 1.55                  & \ifb                  & \multirow{3}{*}{$2011$} & \multirow{3}{*}{\cite{atlasNoteZ}} \\
                              &                           & \muh         & 5184 & 793 & 0.91 && 0.01 && 0.09 && 0.03 & nb     & 1.55                  & \ifb                  &                         &                                    \\
                              &                           & \eh          & 2600 & 449 & 1.00 && 0.02 && 0.13 && 0.04 & nb     & 1.34                  & \ifb                  &                         &                                    \\
                              & \w \to                    & \tau_\had\tn & 2335 & 411 & 11.1 && 0.3  && 1.7  && 0.4  & nb     & 34                    & \ipb                  & 2010                    & \cite{atlasPaperW}                 \\
      \midrule 
      % CMS results.
      \multirow{5}{*}{\cms}   & \multirow{4}{*}{$\z \to$} & \mumu        & 58   & 23  & 1.14 && 0.27 && 0.04 && 0.05 & nb     & \multirow{4}{*}{$36$} & \multirow{4}{*}{\ipb} & \multirow{4}{*}{$2010$} & \multirow{4}{*}{\cite{cmsPaperZ}}  \\
                              &                           & \mue         & 101  & 14  & 0.99 && 0.12 && 0.06 && 0.04 & nb     &                       &                       &                         &                                    \\
                              &                           & \muh         & 517  & 228 & 0.83 && 0.07 && 0.19 && 0.03 & nb     &                       &                       &                         &                                    \\
                              &                           & \eh          & 540  & 346 & 0.94 && 0.11 && 0.22 && 0.04 & nb     &                       &                       &                         &                                    \\
                              & \w \to                    & \tau_\had\tn & 372  & 155 & \multicolumn{8}{c|}{\it not measured} & 18                    & \ipb                  & 2010                    & \cite{cmsNoteW}                    \\
      \midrule
      % LHCb results.
      \multirow{5}{*}{\lhcb}  & \multirow{5}{*}{$\z \to$} & \mumu        & 124  & 42  & 77.4 && 10.4 && 8.6  && 2.7  & pb     & 1.03                  & \ifb                  & \multirow{5}{*}{$2011$} & \multirow{5}{*}{\cite{lhcbPaperZ}} \\
                              &                           & \mue         & 421  & 130 & 75.2 && 5.4  && 4.1  && 2.6  & pb     & 1.03                  & \ifb                  &                         &                                    \\
                              &                           & \emu         & 155  & 57  & 64.2 && 8.2  && 4.9  && 2.2  & pb     & 0.96                  & \ifb                  &                         &                                    \\
                              &                           & \muh         & 189  & 53  & 68.3 && 7.0  && 2.6  && 2.4  & pb     & 1.03                  & \ifb                  &                         &                                    \\
                              &                           & \eh          & 101  & 37  & 77.9 && 12.2 && 6.1  && 2.7  & pb     & 0.96                  & \ifb                  &                         &                                    \\
      \bottomrule
    \end{tabular}
    \caption[something]{A complete summary of the individual $\z \to
      \ditau$ and $\w \to \tau\tn$ cross-section analyses of \atlas,
      \cms, and LHCb using $2010$ and $2011$ datasets. The experiment,
      final state, number of observed events ($N_\mathrm{obs}$),
      number of background events ($N_\mathrm{bkg}$), cross-section
      measurement ($\sigma$), integrated luminosity (\lum), dataset,
      and reference are given for each analysis.\label{tab:results}}
  \end{center}
\end{table*}

\section{$\z \to \ditau$}

Four final states produced from the \tl decays of $\z \to \ditau$
events are considered by all three experiments: two muons (\mumu), a
muon and an electron (\mue), a muon and a hadronic jet (\muh), and an
electron and a hadronic jet (\eh). Six backgrounds to these final
states are considered: Drell-Yan production of di-muon or di-electron
pairs, $WW$ decays, $t\bar{t}$ decays, $Z$ production with an
associated jet, $W$ production with an associated jet, and QCD
multijet events.
 
\subsection{Particle Selection}

All final states are triggered by a muon with a \pt greater than $9-15
~\gev$, except the \eh final state, which is triggered by an electron
with \et greater than $12-15 ~\gev$ for \atlas and \cms and \pt
greater than $10 ~\gev$ for \lhcb. An additional hadronic trigger is
included by \atlas for the \eh final state, while the \lhcb \mue final
state is split into a muon triggered final state (\mue), and an
electron and muon triggered final state (\emu).

Muons are identified by requiring an isolated track associated with
muon system hits. Electrons are identified by requiring an isolated
track associated with electromagnetic calorimeter energy. Both one and
three-pronged hadronic \tl decays are identified within \atlas and
\cms using the anti-\kt jet algorithm and requiring one or three
charged particles, whereas only single-pronged hadronic \tl decays are
identified within \lhcb by requiring a single isolated track with an
associated hadronic calorimeter energy and minimal electromagnetic
calorimeter energy.

\subsection{Event Selection}

The $\z \to \ditau$ signal produces a high mass back-to-back final
state in the transverse plane with missing energy (\met) and a \pt
imbalance between the two \tl decay products due to unreconstructed
neutrinos. Additionally, the lifetime of the \tl produces decay
products with an enhanced impact parameter. Selection requirements
based on these five signatures are used by the three experiments to
separate signal from background.

For \atlas, a visible mass selection requirement is placed on all four
final states, and a transverse mass requirement on the \muh and \eh
final states. A requirement on the transverse separation of the \tl
decay products and the \met of the event is used for all final states
except \mumu, where further requirements on the angular separation,
\pt asymmetry, and impact parameters of the two \tl decay products are
applied.

For \cms, only a transverse mass requirement is applied to the \mue,
\muh, and \eh final states. Due to the large Drell-Yan background to
the \mumu final state, requirements are placed on the visible mass,
transverse separation between the muons and \met, the muon \pt
asymmetry, and the impact parameter of the muons.

Unlike for \atlas and \cms, missing energy cannot be measured within
\lhcb. However, a high resolution vertex locator allows for strict
requirements to be placed on the \tl decay product impact parameters
for the \mumu, \muh, and \eh final states. Both a visible mass and
transverse separation requirement is placed on all final states, while
an additional \pt asymmetry requirement is also placed on the \mumu
final state.

\subsection{Background Estimation}

Drell-Yan production of lepton pairs is the primary background to the
\mumu final state for all three experiments, as well as the \eh final
state for \atlas and \cms. The Drell-Yan visible mass shape is
determined for \atlas from simulation, while for both \cms and \lhcb
the template is obtained with a reversed impact parameter
requirement. For \atlas and \lhcb the template is normalized to the
on-shell \z mass peak, and for \cms normalized to an impact parameter
side-band.

The QCD multijet background is large in the \muh and \eh final states,
and a visible mass shape is determined from data for all three
experiments by requiring candidates with same-sign charge. The
normalization is also taken from data, scaling the number of same-sign
events by the estimated opposite-sign/same-sign event ratio for the
background.

The \w with jets background mass shape is determined from simulation
for all three experiments and normalized using transverse mass
side-bands for \atlas and \cms, and a same-sign side-band for
\lhcb. The \z with jets visible mass shape is taken from simulation
for \atlas and \lhcb, and from a reversed impact parameter requirement
for \cms. The background is normalized using a visible mass side-band
for \atlas, an impact parameter side-band for \cms, and a same-sign
side-band for \lhcb.

For all three experiments the visible mass distributions for the $WW$
and $t\bar{t}$ backgrounds are estimated from simulation. The
normalization for these backgrounds is also taken from simulation for
\atlas and \lhcb, and from a transverse mass side-band for \cms. Both
the $WW$ and $t\bar{t}$ background contributions are minimal for all
final states.

\subsection{Systematics}

For both \atlas and \cms the hadronic \tl identification efficiency
and energy scale is the primary systematic uncertainty for the \muh
and \eh final states, ranging from $8\%-23\%$. In the \mumu final
state the primary uncertainty is between $2\%-9\%$ from muon
efficiency and acceptance, and for the \mue final state is between
$2\% - 6\%$ from electron efficiency.

For \lhcb, the Drell-Yan background provides the largest systematic
uncertainty of $8\%$ to the \mumu final state. Electron reconstruction
efficiency contributes the primary uncertainty to the \mue, \emu, and
\eh final states of $4\%$, while the impact parameter selection
efficiency provides a $2\%$ uncertainty to the \muh final state.

\subsection{Results}

The measured cross-section, number of observed events, and number of
background events for each final state of all three
experiments is given in Table~\ref{tab:results}. Note the reduced
statistics of the \lhcb results, due to the acceptance of the
detector, but the enhanced purity of the \mumu final state.

The combined $\z \to \ditau$ cross-section measurement for each
experiment is given in Table~\ref{tab:combined}, including the
fiducial definition and predicted theory result. The \atlas and \cms
theory predictions were calculated using \fewz~\cite{fewz}, while the
\lhcb prediction was calculated with \dynnlo~\cite{dynnlo}. The \atlas
combined result does not include the \mumu final state.

\begin{table}
  \small
  \begin{center}
    \settowidth{\length}{\space}
    \begin{tabular}{c|RERERER@{\extracolsep{\length}}l@{\extracolsep{-2pt}}|RER@{\extracolsep{\length}}l}
      \toprule
      \multicolumn{1}{c}{Exp.} & 
      \multicolumn{8}{c}{$\sigma\pm\mathrm{stat.}\pm\mathrm{syst.}\pm\mathrm{lumi.}$} & 
      \multicolumn{4}{c}{$\sigma$ theory} \\
      \midrule[\heavyrulewidth]
      % ATLAS results.
      \multirow{2}{*}{\atlas} & 0.92 && 0.02 && 0.08 && 0.03 & \nb                               & 0.96 && 0.05 & \nb                           \\
                              & \multicolumn{8}{c|}{$66 < M_\ditau < 116 ~\gev$}                  & \multicolumn{4}{c}{\fewz}                    \\
      \midrule
      % CMS results.
      \multirow{2}{*}{\cms}   & 1.00 && 0.05 && 0.08 && 0.04 & \nb                               & 0.97 && 0.04 & \nb                           \\
                              & \multicolumn{8}{c|}{$60 < M_\ditau < 120 ~\gev$}                  & \multicolumn{4}{c}{\fewz}                    \\
      \midrule
      % LHCb results.
      \multirow{3}{*}{\lhcb}  & 71.4 && 3.5  && 2.8  && 2.5 & \pb                                & 74.3 && 2.1  & \pb                           \\
                              & \multicolumn{8}{c|}{$\pt^\tau > 20 ~\gev, 2.0 < \eta^\tau < 4.5$} & \multicolumn{4}{c}{\multirow{2}{*}{\dynnlo}} \\
                              & \multicolumn{8}{c|}{$60 < M_\ditau < 120 ~\gev$}                  &                                              \\ 
      \bottomrule
    \end{tabular}
    \caption{Combined $\z \to \ditau$ and theoretical cross-section
      results for the three experiments. The \atlas result does not
      include the \mumu final state.\label{tab:combined}}
  \end{center}
\end{table}

\section{$\w \to \tau\tn$}

Due to the large $\w \to \lep\lepn$ background to leptonically
decaying \tls produced from \w bosons, only hadronic decays of the \tl
are considered for the \atlas and \cms $\w \to \tau\tn$ analyses. QCD
jets, $\w \to \lep\lepn$, $\w \to \tau_\lep\tn$, and \z with jets
events provide the primary backgrounds to this signal. Of the three
experiments, only \atlas has performed a $\w \to \tau\tn$
measurement. An observation has been made with \cms, but without a
cross-section measurement.

\subsection{Particle Selection}

A $\pt > 12 ~\gev$ trigger on the hadronic \tl combined with a $\met >
20 ~\gev$ trigger is used to select events for \atlas, while a $\pt >
20 ~\gev$ hadronic \tl trigger and $\met > 25 ~\gev$ trigger is used
for \cms.

Hadronic \tls are selected for \atlas using a boosted decision tree
based on the collimation, impact parameter, and lead track \pt over
electromagnetic calorimeter energy of the hadronic \tl candidate. For
the \cms selection, requirements are placed on the hadronic \tl lead
track \pt, isolation, and associated muon hits.

\subsection{Event Selection}

To eliminate \w and \z with jets backgrounds, events with high \pt
leptons outside the hadronic \tl jet are rejected for both \atlas and
\cms. A large \met is also required for both to reduce the QCD jet
background. Additionally, a high \met significance and transverse
separation between the \tl and \met is required for \atlas, while the
ratio of the \tl jet \pt to the \pt of the remaining jets is required
to be large for \cms.

\subsection{Background Estimation}

The \w and \z background transverse mass shapes for both \atlas and
\cms are determined and normalized using simulation. The QCD jet
background transverse mass shape and normalization is determined using
an $ABCD$ method for both. For \atlas, requirements are made on the
\met significance and hadronic \tl identification, while requirements
on the \met and ratio of \tl jet \pt to the \pt of the remaining jets
are used for \cms.

\subsection{Systematics}

The primary systematic uncertainty for the \atlas $\w \to
\tau_\had\tn$ cross-section measurement is due to the identification
efficiency for the hadronic \tl jets and estimated to be $10\%$. For
\cms, no cross-section measurement was made, and so no uncertainty
analysis is available.

\subsection{Results}

The number of observed $\w \to \tau\tn$ events and background events
is given in Table~\ref{tab:results} for both \atlas and \cms, as well
as the measured \atlas cross-section. To determine the total
cross-section, the \atlas $W \to \tau_\had\tn$ cross-section is
extrapolated from the measured fiducial region to full acceptance, and
divided by the experimentally known branching fraction for \tls to
hadrons. The result given in Table~\ref{tab:results} agrees well with
the theoretically predicted value of $10.5 \pm 0.5 ~\nb$ calculated
using \fewz.

\section{Conclusion}

\begin{figure}[!htb]
  \begin{center}
    \ifdraft
    {\includesvg[width=\columnwidth,name=fig1,pretex=\footnotesize]{comparison}}
    {\includegraphics[width=\columnwidth]{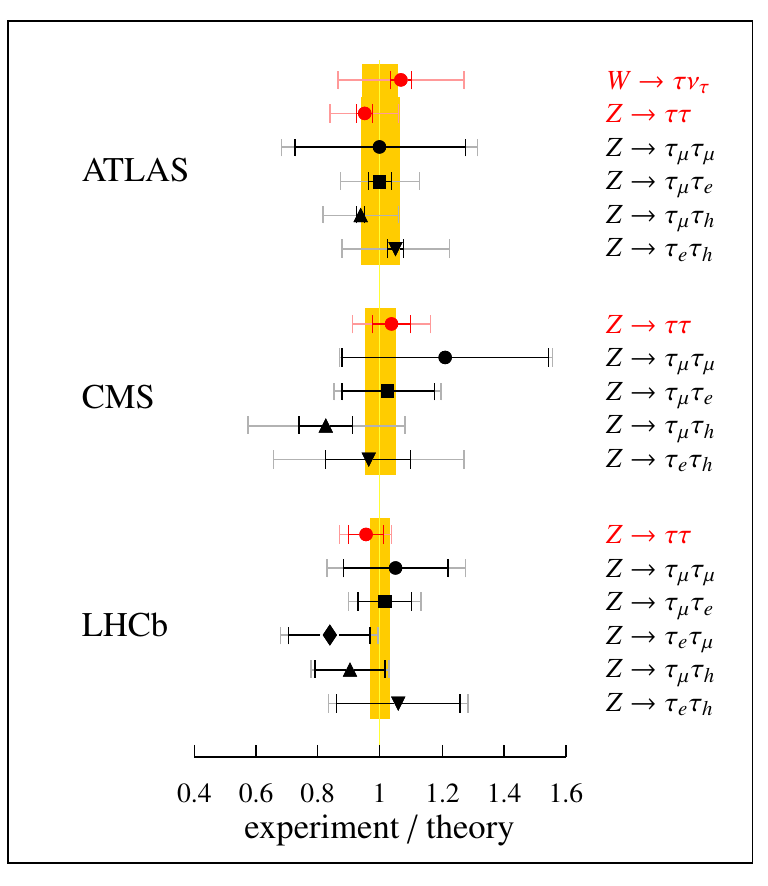}}
  \end{center}
  \caption{Ratios of the experimental cross-section measurements of
    Table~\ref{tab:results} and Table~\ref{tab:combined} to their
    expected theoretical values. The combined results are given in red
    and the individual final states in black. The dark error bars are
    the statistical uncertainty, while the light error bars are the
    combined systematic uncertainty and uncertainty due to the
    integrated luminosity. The dark yellow band indicates the
    theoretical uncertainty centered about the light yellow
    line.\label{fig:comparison}}
\end{figure}

A comparison of the measured cross-sections of Table~\ref{tab:results}
and Table~\ref{tab:combined} divided by their predicted theoretical
values is given in Figure~\ref{fig:comparison}. The red points
represent combined results and the black points individual final
states. The dark error bars correspond to statistical uncertainty,
while the light error bars correspond the combined systematic
uncertainty and uncertainty due to the integrated luminosity. The
light yellow line represents a ratio of unity, while the dark yellow
band indicates the theoretical uncertainty of the prediction.

There is good agreement between all measured cross-sections and their
theoretical values. Currently, \atlas provides the most statistics for
the combined $\z \to \ditau$ cross-section measurement, while \lhcb
yields the most precise cross-section measurement. The \atlas $\w \to
\tau\tn$ measurement has a large systematic uncertainty and does not
allow for a more precise test of the $\w \to \tau\tn$ to $\w \to
\mu\mn$ ratio measured at LEP.

\bibliographystyle{elsarticle-num} 
\bibliography{proceedings}
\end{document}